\documentclass{article}
\usepackage[version=4]{mhchem}
\usepackage{graphicx} 
\usepackage{authblk}
\usepackage{blindtext}
\usepackage{titlesec}
\usepackage{wrapfig}
\usepackage[subrefformat=parens]{subcaption}
\usepackage[font=footnotesize, labelfont=bf]{caption}
\usepackage[acronym]{glossaries}
\usepackage[usetitle]{rsc}
\usepackage{siunitx}
\usepackage[a4paper, total={6in, 8in}, margin=0.5in]{geometry}
\usepackage[dvipsnames]{xcolor}
\usepackage{float}
\usepackage{subfig}
\usepackage{multirow}
\usepackage{physics}
\usepackage{url}
\title{Solphin: Photovoltaic efficiency analysis for bulk materials using Python}
\author[1]{Philippa U. Cox}
\author[1]{Peter P. Russell}
\author[2]{Andrea Crovetto}
\author[1]{Alexander G. Squires}
\author[1]{Louie Slocombe}
\author[1]{David O. Scanlon}
\affil[1]{School of Chemistry, University of Birmingham, Edgbaston, Birmingham, B15 2TT, UK}
\affil[2]{DTU Nanolab, Technical University of Denmark, DK-2800 Kgs. Lyngby, Denmark}
\date{}

\begin{document}
\maketitle

\section*{Summary}

Photovoltaic materials convert light into electricity through the photovoltaic effect. These materials are the fundamental building blocks of solar cells, devices engineered to maximise light capture, convert it to electricity, and deliver it efficiently to the load. Solar cells are becoming increasingly common in day-to-day life, with solar farms being built across the countryside and in industrial complexes, as well as for household rooftop electricity generation. As the world continues to require urgent solutions to reduce carbon emissions, solar technology has emerged as one of the most promising, cheap, and sustainable methods for addressing the current climate crisis \cite{malik2025comprehensive}. Although researchers have achieved great success in developing high-efficiency photovoltaic materials, many of the highest-performing inorganic materials contain highly toxic or expensive elements. Therefore, it is crucial that research is continued in order to simultaneously improve the efficiency while also reducing the environmental and economic impact of solar cell devices\cite{kirchartz2025state}. 

One of the key stages in this area of research is the computational screening of novel materials in order to increase the speed of the materials discovery process and reduce the cost of laboratory analysis through the removal of inefficient candidate materials. The complete packages for modelling the efficiency of photovoltaics tend to focus on the device level and not the properties of the bulk photovoltaic material, for example PV LIB \cite{holmgren2018pvlib}, OTSun \cite{cardona2020otsun}, PVAnalytics \cite{osti_1887283}, PYPVCell\cite{lee2017pypvcell} and Solcore\cite{alonso2018solcore}. Currently, analysing bulk materials for photovoltaic applications is complex, with multiple analytical methods split across various packages with varying current levels of support and maintenance. These include SL3ME\cite{Williams2013} and the SLME functionality of Pymatgen\cite{ong2013python} for calculating efficiency metrics, and effmass\cite{Whalley2018} for calculating carrier effective masses. Consequently, the reporting of the resulting efficiency in the literature is often inconsistent, especially for theoretical studies. Solphin provides a comprehensive package that takes users from initial structural input to a full analysis of photovoltaic properties, including the novel photovoltaic figure of merit published by Crovetto.\cite{crovetto2024phenomenological,Crovetto2026} Documentation guides users through Density Functional Theory (DFT) calculations from initial relaxation to optical and electronic structure simulations.

Solphin provides these key features:

\begin{itemize}
    \item Integration of existing photovoltaic efficiency metrics and their dependent properties into a single package.
    
    \item Inclusion of a novel, less idealized, figure of merit that accounts for a range of optical and defect properties of photovoltaic materials \cite{crovetto2024phenomenological,Crovetto2026}.

    \item For quantities straightforward to obtain from ab-initio calculations, Solphin offers automatic input generation and output processing for the commonly used VASP\cite{Kresse1994,Kresse1996a,Kresse1996b} and CASTEP\cite{Clark2005} codes. This allows for reproducible and efficient calculations and analysis.
    
    \item For the more complex quantities, interactive sliding scale functionality helps illustrate the impacts of doping density, non-radiative recombination lifetime, and carrier mobility on PV efficiency.
    
    \item Post-processing tools, including plotting, which allow for high-quality, publication-ready plots and outputs. 
    
    \item Flexible implementation of the above methods to enable the determination of features with minimal necessary input. For example, the SQ limit requires only a band gap, and the SLME just an additional optical absorption simulation.

    \item Support for a range of illuminance spectra with the SQ and Blank metrics, including the AM1.5 outdoor spectrum, the CIE 15:2018 LED spectra, pure R/G/B LED spectra, and fluorescent and IR spectra.
        
\end{itemize}

\section*{Solphin}

\begin{figure}[H]
    \centering
    \includegraphics[width=0.95\linewidth]{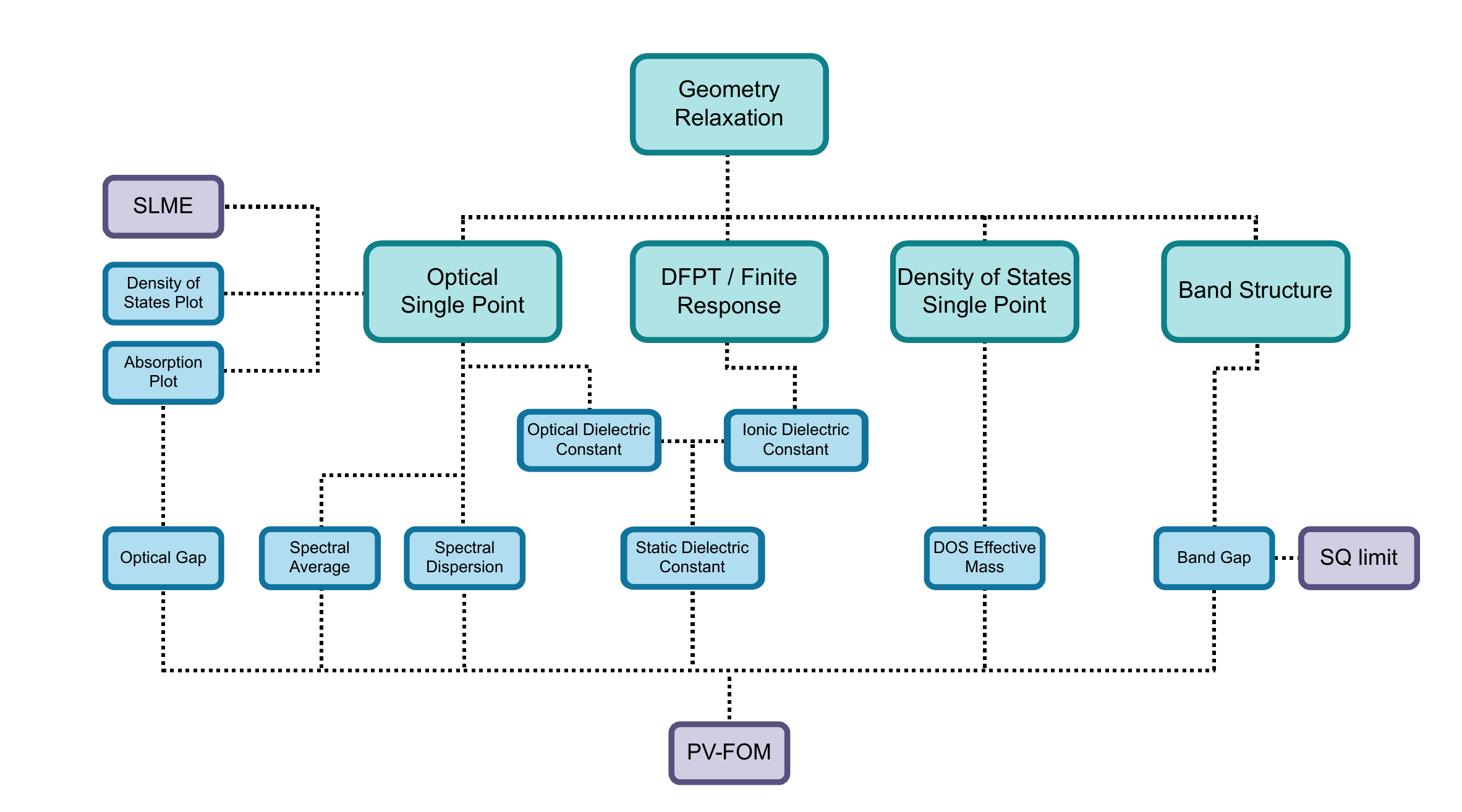}
    \caption{The basic workflow structure, with DFT calculations given in teal, primary outputs given in purple and secondary outputs given in blue.}
    \label{fig:workflow}
\end{figure}

Solphin is a Python package that provides facilities to analyse the photovoltaic potential of an inorganic bulk material from an initial structure and/or experimentally determined properties. To do so, Solphin provides a workflow (Figure \ref{fig:workflow}) for inorganic photovoltaic material characterisation, improving the reproducibility and efficiency of calculations and enabling direct comparison between materials. 

The full functionality generates VASP or CASTEP input files to run the required DFT calculations to assess photovoltaic efficiency. The automatic input functionality supports a range of DFT functionals: LDA, GGA functionals PBEsol and PBE, meta-GGA functional R2SCAN and the hybrid functionals HSE06 and PBE0. It has the capability to include dispersion corrections, as well as any user declared input flags. When calculating final results, Solphin allows manual input for cases where values may already be known through external calculations or experiment. The outputs generated are:

\begin{itemize}
    \item Shockley-Queisser detailed balance limit plots \cite{shockley1961detailed}.
    \item The Spectroscopic Limited Maximum Efficiency (SLME) and the Blank metric for thin film efficiency \cite{yu2012identification, blank2017selection}.
    \item Band structure and density of states plots with band gap output. 
    \item Density of states effective mass, with band edge and charge carrier analysis.
    \item Optical absorption onset.
    \item Static dielectric constant.
    \item The spectral average and dispersion of the absorption coefficient. 
    \item The photovoltaic figure of merit (PV-FOM) \cite{crovetto2024phenomenological}.
\end{itemize}

The generation of the band structure calculations, the analysis and plotting of the density of states, band structure, and absorption were adapted and wrapped from sumo \cite{ganose2018sumo}. The detailed balance code was inspired by the detailed balance code by Kaklin \textit{et al.} \cite{kaklin2024sqlimit, byrnes_solarcellefficiencylimits}. The SLME metric is wrapped from pymatgen.\cite{ong2013python} The main dependencies are matplotlib for plotting, numpy for calculation, scipy for scientific constants, and pymatgen for structure interaction and input generation \cite{Hunter:2007,harris2020array, 2020SciPy-NMeth, ong2013python}.

\section*{Future Development}

Currently, Solphin input generation and output analysis are limited to VASP and CASTEP; future development aims to extend support to CP2K and Quantum Espresso. This version treats carrier lifetimes, carrier mobilities, and dopant density as variables, but future integration with AMSET\cite{Ganose2021} and carrier capture will allow these to be calculated directly. 

\section*{Software Availability}

The full repository is distributed under the open-source MIT licence and is available at \url{https://github.com/SMTG-Bham/solphin}, as well as through PyPI under \texttt{solphin}.

\section*{Author Contributions}

P.U.C.: Developed the idea for Solphin, researched and wrote the code for the mathematical and physics-based functions, software development and debugging, testing, writing and documenting the methodologies. P.P.R.: Software development including managing standardisation and adherence to coding standards, development of the visualisation components as well as the package management and documentation. L.S.: Software development including CASTEP integration, standardisation, debugging, and packaging. A.C.: Developed the photovoltaic figure of merit and the initial base code for the calculation of this metric. Provided valuable advice on implementation. A.G.S.: Supervision and advice on best code development practice, provided the base code for the calculation inputs. D.O.S.: Supervision, management of resources and funding. 

\section*{Conflicts of Interest}

There are no conflicts to declare. 

\section*{Acknowledgements}

The authors would like to extend a huge thank you to Brooke Busbee for creating the branding for Solphin. P.U.C, P.P.R, A.G.S, L.S and D.O.S acknowledge the use of the University of Birmingham's BlueBEAR HPC service (\url{https://www.birmingham.ac.uk/bear}), the Baskerville Tier 2 HPC service (https://www.baskerville.ac.uk/; funded by the EPSRC and UKRI through the World Class Labs scheme (EP/T022221/1) and the Digital Research Infrastructure programme (EP/W032244/1)), and the Sulis Tier 2 HPC platform hosted by the Scientific Computing Research Technology Platform at the University of Warwick (funded by EPSRC Grant EP/T022108/1 and the HPC Midlands+ consortium). Through our membership of the UK's HEC Materials Chemistry Consortium, which is funded by the UK Engineering and Physical Sciences Research Council (EPSRC; EP/L000202, EP/R029431, EP/T022213), this work also used ARCHER2 UK National Supercomputing Services. We are also grateful to the UK Materials and Molecular Modelling Hub for computational resources, which is partially funded by EPSRC (EP/T022213/1, EP/W032260/1 and EP/P020194/1). P.P.R and D.O.S. gratefully acknowledge the PRAETORIAN project, funded by UK Research and Innovation (UKRI) under the UK government's Horizon Europe funding guarantee (EP/Y019504/1).

We thank Se\'an R. Kavanagh for providing CdTe data. We thank Jacob Baggott for his valuable input in the office, and all members of the Scanlon Materials Theory Group for their useful feedback and discussion. 

\section*{AI Usage Disclosure}

Claude (Sonnet 4.6 and Fable 5) was used to assist with debugging and CASTEP integration, and to assist with the dos\_eff\_mass module by helping develop ideas for implementing the physics. AI tools were not used to generate the underlying physics of the code. ChatGPT (5.3 mini) was used to generate doc-strings based on a user given template. The authors extensively reviewed, validated, and edited any AI-assisted work. AI was not used in the writing of this manuscript, nor in the design and architecture choices of the software. 

\bibliography{ref}

\end{document}